\documentclass[twocolumn,amsmath,amssymb,longbibliography,aps,prl,reprint,
superscriptaddress]{revtex4-1}
\usepackage{graphicx}
\usepackage{amsmath}
\usepackage{bm}
\usepackage{comment}
\usepackage{multirow}
\usepackage{color}
\usepackage{amsfonts}
\usepackage{booktabs}
\usepackage{ulem}
\usepackage{siunitx}

\newcommand{\red}[1]{{ #1}}

\begin{document}

\title{Tunable Mixed Parity Spin Splittings in Altermagnets}

\author{Yue Yu}
\affiliation{School of Physics and Astronomy and William I. Fine Theoretical Physics Institute, University of Minnesota, Minneapolis, Minnesota 55455, USA} 

\begin{abstract}
Momentum-dependent spin splitting and its relation to inversion ($P$) and time-reversal ($T$) symmetries are central to nonrelativistic spintronics. Representative examples include collinear altermagnets with $(P,T)=(+,-)$ and non-collinear odd-parity magnets with $(P,T)=(-,+)$. In this work, we develop a theoretical framework to induce odd-parity spin splittings in the more abundant collinear altermagnets through two mechanisms: driving by a two-color linearly polarized light field or coupling to a $P$-odd loop-current order. Properly phase-locked two-color driving induces a static $(P,T)=(-,-)$ order, symmetry-equivalent to a translationally invariant $P$-odd loop-current order. Coupling this order to an altermagnet produces a controllable mixed-parity spin texture, opening new avenues for the electrical and optical manipulation of spin-polarized currents in spintronics applications. The same mechanism applied to a collinear $PT$-symmetric magnet induces a distinct $(P,T)=(+,+)$ state with a nonrelativistic dissipationless anomalous spin Hall conductivity. We present group-theory and microscopic Floquet theory to highlight the emergent responses.
\end{abstract}
\maketitle

Altermagnetism has recently emerged as a major frontier in quantum
magnetism\cite{hayami2019,yuan2020,mazin2021,vsmejkal2022,vsmejkal2022emerging,fukaya2025}.
Altermagnets exhibit collinear spin ordering while preserving lattice translation
and inversion ($P$) symmetry, but breaking time-reversal ($T$) symmetry, thereby
sharing key features with ferromagnets.
This $(P,T)=(+,-)$ symmetry gives rise to nonrelativistic even-parity spin splitting in momentum space\cite{krempasky2024}, which underpins spin transport\cite{gonzalez2021,han2024}
and spin caloritronics\cite{cui2023}.
In the presence of spin-orbit coupling (SOC), altermagnets further support anomalous
Hall transport\cite{vsmejkal2020,Roig:2025}.

The emergence of altermagnetism has also stimulated broader interest in unconventional
magnets\cite{jiang2024,xiao2024,chen2024,watanabe2024Symmetry}, with odd-parity
(or $p$-wave) magnets attracting particular
attention\cite{hellenes2023,matsuda2024,brekke2024,yu2025,luo2025,song2025}.
These antiferromagnets(AFMs) host noncollinear spin orderings that break lattice
translational symmetry, yet induce a translationally invariant spin order with
$(P,T)=(-,+)$, leading to odd-parity Bloch spin splitting\cite{hellenes2023,brekke2024,matsuda2024,yu2025,luo2025}.
As nonrelativistic analogues of Rashba SOC, odd-parity magnets hold
significant promise for spintronic applications\cite{manchon2015,gonzalez2024,hu2024,chakraborty2025}.

In nature, collinear magnetic states are more common and stable than non-collinear ones. It is therefore of practical interest to induce odd-parity spin splitting from these more prevalent collinear states, either through external driving fields or via internal coupling to additional symmetry-breaking orders. This strategy has been explored in collinear $PT$-symmetric magnets, which has $(P,T)=(-,-)$. The combined $PT$ symmetry enforces doubly Kramers-degenerate bands\cite{zhang2014,watanabe2020,watanabe2024,hayami2022}. Once time-reversal symmetry is broken, odd-parity spin splitting with $(P,T)=(-,+)$ can emerge in the absence of SOC. This mechanism forms the basis of recent studies on $PT$-symmetric magnets coupled to even-parity loop-current order\cite{lin2026}, as well as Floquet engineering using circularly polarized light\cite{Zhou2025,zhu2025,li2025,huang2025,fu2026,fu2026floquet}.

In this work, we propose a strategy to induce nonrelativistic odd-parity spin splitting in altermagnets. With more than 200 candidate materials\cite{bai2024}, altermagnets provide an even more experimentally accessible platform than $PT$-symmetric magnets. More importantly, because altermagnets intrinsically host even-parity spin splitting, introducing additional odd-parity splitting through external control creates a uniquely tunable platform in which both parities coexist. The resulting mixed-parity spin texture serves as a nonrelativistic analogue of an altermagnet subject to a {\it tunable} odd-parity Ising SOC, opening spintronic functionalities that are inaccessible in either pure phase alone.

Because altermagnets and odd-parity magnets differ in both $P$ and $T$ symmetries, the applied field must break both symmetries to convert one into the other. There are two possible routes to achieve this: (1) breaking $P$, $T$, and their product $PT$ simultaneously, or (2) breaking both $P$ and $T$ while preserving the combined $PT$ symmetry. The first approach can be realized by combining $P$-odd and $T$-odd driving fields. Here, a $P$-odd (and $T$-even) driving field can be gating electric field in thin films\cite{mazin2023}, or internal coupling to P-odd orbital orders \cite{gu2025,Duan2025AFAM,li2026p}. In this work, we focus on the second approach, where the driving field itself satisfies $(P,T)=(-,-)$.

This choice is motivated by the need for an unambiguous response signal. In approach (1), breaking $P$, $T$, and $PT$ in an altermagnet induces not only odd-parity spin splitting, but also an additional $PT$-symmetric magnetic order. Upon introducing dissipation\cite{Freimuth2014,Jungwirth2017,watanabe2024,oiwa2022}, the resulting $P$-odd responses from this unwanted order can obscure those associated with the odd-parity magnetic phase. Furthermore, although SOC is not the focus of this work, it is always present in realistic systems, and approach (1) would lift the Kramers degeneracy even in the nonmagnetic phase. By contrast, preserving $PT$ symmetry in approach (2) guarantees doubly degenerate bands in the nonmagnetic phase, ensuring that the mixed-parity spin splitting arises unambiguously from the interplay between the driving field and the altermagnetic order.

\begin{figure}[h]
\centering
\includegraphics[width=0.95\linewidth]{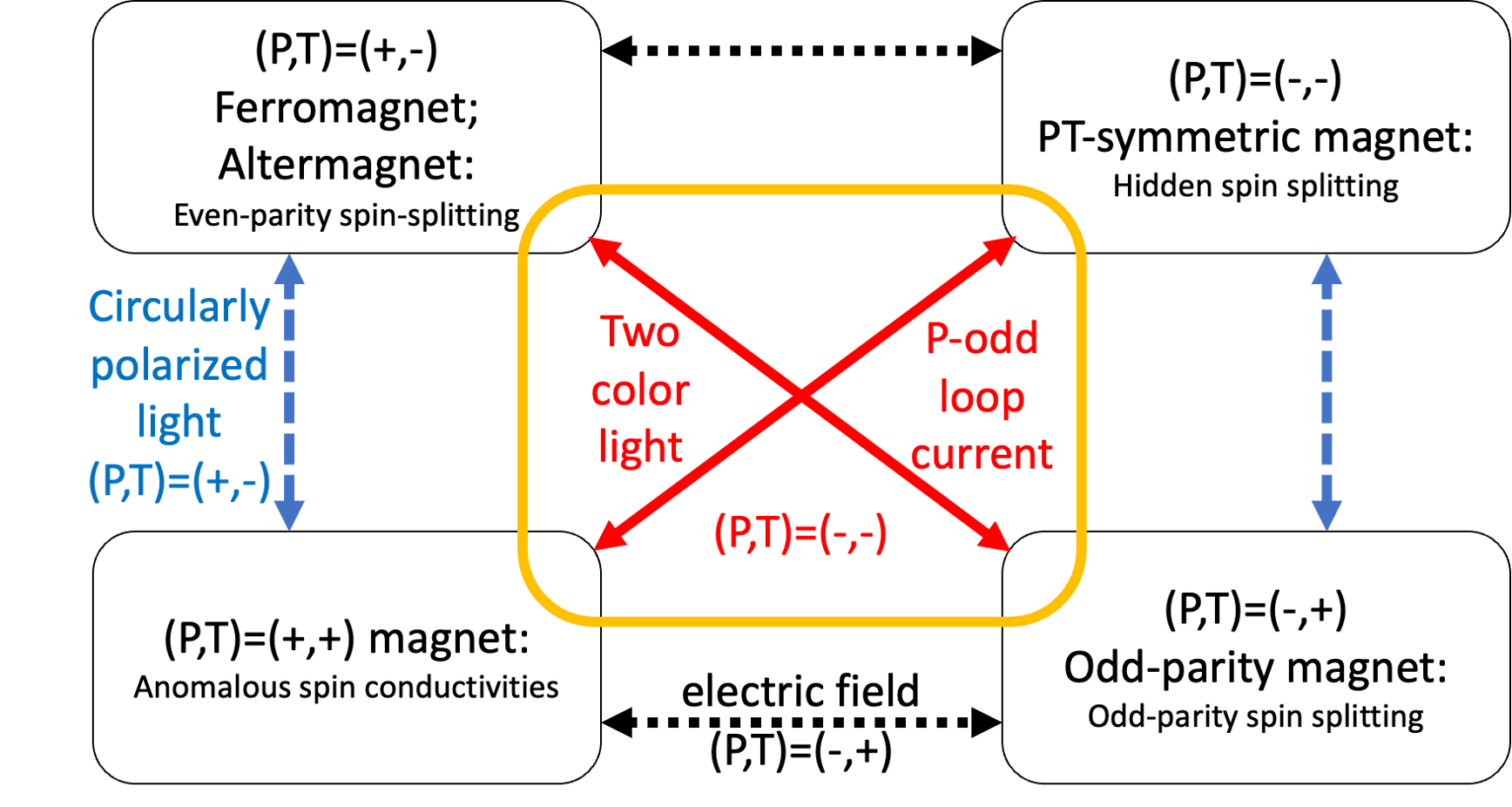}
\caption{\red{Schematic picture of four classes of magnetic states classified by $(P,T)$ symmetry and controlling fields connecting them. Gating electric field (horizontal dotted lines) breaks inversion symmetry while preserving time-reversal symmetry. Circularly polarized light (vertical dashed lines) breaks T symmetry while preserving P symmetry. This work is about the two-color linearly polarized light field (or coupling to a P-odd loop-current order), which breaks both P and T symmetry while preserving the product (diagonal solid lines). When coupled to altermagnets, the two-color light leads to nonrelativistic mixed-parity spin splitting; when coupled to collinear $PT$-symmetric magnets, they generate nonrelativistic dissipationless transverse spin conductivity.}}
\label{F:1}
\end{figure}

We use group-theoretical analysis to demonstrate that a two-color linearly polarized light field, with an appropriate phase locking between the fundamental and second-harmonic components, satisfies $(P,T)=(-,-)$ and therefore realizes approach (2). Two-color driving is a readily accessible experimental control through second-harmonic generation, a standard technique widely employed in laser spectroscopy\cite{franken1961,shen1989,boyd2008}. We show that the effect of such a driving field is symmetry-equivalent to an internal coupling to a $P$-odd loop-current order, such as the $\Theta_{II}$ loop-current proposed in cuprates\cite{varma2006} and the induced ${\bf Q}=0$ order originating from $P$-odd triple-${\bf Q}$ loop currents in kagome metals AV$_3$Sb$_5$\cite{christensen2022}. We then apply Landau theory to demonstrate the emergence of odd-parity spin splitting in an altermagnet. Furthermore, we construct a two-sublattice microscopic Floquet model for altermagnets coupled to the light field and verify the group-theoretical predictions through both numerical and analytical calculations.

Coupling the same $(P,T)=(-,-)$ order to a $PT$-symmetric magnet induces a distinct class of magnetic states with $(P,T)=(+,+)$. In the absence of the driving field, such a translationally invariant state can arise as a composite order in $P$-even coplanar antiferromagnets\cite{yu2026}, analogous to the emergence of odd-parity magnets in $P$-odd coplanar AFMs. This $(P,T)=(+,+)$ phase breaks spin-rotational symmetry and serves as a nonrelativistic analogue of $P$-even $\vec{L}\cdot\vec{S}$ SOC. Although the bands remain doubly degenerate due to the preserved $PT$ symmetry, spin responses such as the dissipationless anomalous spin Hall effect become allowed.

\red{Figure 1 provides a schematic overview distinguishing between different control methods. A gating electric field \cite{mazin2023}, or internal coupling to odd-parity charge orders \cite{gu2025,Duan2025AFAM,li2026p}, breaks inversion symmetry while preserving time-reversal symmetry (horizontal dotted lines). When applied to a PT-symmetric magnet with Kramers-degenerate bands, this can unlock ferromagnetic or altermagnetic spin splittings. A circularly polarized light field \cite{Zhou2025,zhu2025,li2025,huang2025,fu2026,fu2026floquet}, or internal coupling to even-parity current-loop orders \cite{lin2026}, breaks time-reversal symmetry while preserving inversion symmetry (vertical dashed lines). When applied to a PT-symmetric magnet, this can unlock odd-parity spin splittings. In these two cases, the resulting spin splittings are either odd-parity or even-parity. A two-color light field can break both inversion and time-reversal symmetry while preserving their product (diagonal solid lines). It can induce odd-parity spin splitting in a ferromagnet or altermagnet, providing a tunable mixed-parity spin texture. }

{\it Group theory: }
The linearly polarized two-color light field is described by the vector potential
${\bf A}=(A_1\cos 2\omega t,\, A_2\cos(\omega t+\phi),0)$.
Because the two components have different frequencies, AC responses can arise from the time-dependent composite order $A_xA_y$\cite{yarmohammadi2026}. In contrast, DC responses, including symmetry breaking in the band structure, are governed by time-independent composite orders. Since ${\bf A}$ is odd under both $P$ and $T$, static combinations such as $\overline{A_x^2}$ or $\overline{A_y^2}$ preserve both symmetries, where the overline denotes time averaging.

Owing to the 2:1 frequency ratio, the lowest-order nontrivial static composite orders contain one factor of $A_x$ and two factors of $A_y$. Examples include $\overline{A_xA_y^2}$, $\overline{A_xA_y\partial_t A_y}$, $\overline{A_y^2\partial_t A_x}$, and terms involving higher-order time derivatives (overline denotes time-average). Their $(P,T)$ properties become transparent after time averaging. The leading contribution, together with even-order time-derivative terms, satisfies
$\overline{A_xA_y^2}\propto\cos 2\phi$
and transforms as $(P,T)=(-,-)$. By contrast, the subleading contributions, together with other odd-order time-derivative terms, satisfy
$\overline{A_xA_y\partial_t A_y},\,\overline{A_y^2\partial_t A_x}\propto\sin 2\phi$
and transform as $(P,T)=(-,+)$. To realize a purely $(P,T)=(-,-)$ driving field while preserving $PT$ symmetry, the relative phase should be chosen as $\phi=0$ or $\pi/2$. In the following analysis, we take $\phi=0$.

Other real-space symmetries can be analyzed within Landau theory (we follow the notation of Ref.~\cite{McClarty2023} and neglect SOC). The vector potential transforms as
$A_i\in\Gamma_V\otimes\Gamma_1^S$,
where $\{x,y,z\}\in\Gamma_V$ denotes the real-space polar vector representation and $\Gamma_1^S$ denotes a spin scalar. The composite order
$L_A\equiv\overline{A_xA_y^2}$
transforms as
$\Gamma_L\otimes\Gamma_1^S\subset\Gamma_V\otimes\Gamma_V\otimes\Gamma_V\otimes\Gamma_1^S$.
The altermagnetic order $\vec{N}$ transforms as
$\Gamma_N\otimes\Gamma_A^S$,
where $\Gamma_N$ is determined by the site symmetry responsible for the $k$-even spin splitting, and $\Gamma_A^S$ is the vector representation under spin rotations. The induced odd-parity spin splitting $\vec{O}$ transforms as
$\Gamma_O\otimes\Gamma_A^S\subset(\Gamma_L\otimes\Gamma_1^S)\otimes(\Gamma_N\otimes\Gamma^S_A)=\Gamma_L\otimes\Gamma_N\otimes\Gamma_A^S$,
and arises through the Landau coupling term
$(\vec{O}\cdot\vec{N})L_A$.

As an example, in tetragonal $D_{4h}$ systems, the above vector potential gives
$\Gamma_L\sim k_xk_y^2\in E_u$.
For an altermagnet with $k_xk_y\in B_{2g}$ spin splitting, the light-induced odd-parity spin splitting follows
$\Gamma_O\sim k_y\in B_{2g}\otimes E_u=E_u$.
Its spin orientation follows that of the altermagnetic order: $\vec{O}\parallel\vec{N}$.

The $(P,T)=(-,-)$ composite order $L_A$ carries the same symmetry as a translationally invariant odd-parity loop-current order $L_C$. Consequently, the two-color light field can induce such loop-current orders through the Landau coupling term $L_AL_C$. In the absence of light, internal coupling between $P$-odd loop-current orders and altermagnets can likewise generate odd-parity spin splitting through the Landau coupling
$(\vec{O}\cdot\vec{N})L_C$.

{\it Microscopic models:}
To verify the above analysis, we study a microscopic two-sublattice model \cite{antonenko2025,roig2024}:
\begin{equation}
\begin{split}
h &= t_{x,{\bf k}}\tau_x + t_{z,{\bf k}}\tau_z + J\tau_z\sigma_z
     + \varepsilon_{0,{\bf k}}, \\
E_{\alpha\beta} &= \varepsilon_0 + \alpha\sqrt{t_x^2 + (t_z + \beta J)^2},
\quad \alpha,\beta = \pm,
\end{split}
\label{E:1}
\end{equation}
where $\tau$ and $\sigma$ are Pauli matrices acting in the sublattice and spin spaces, respectively, and spin-orbit coupling is neglected. Time-reversal and inversion symmetries are represented by
$T = i\sigma_y K$
and
$P = \mathbf{1}$
together with momentum inversion. The hopping parameters $t_{x}$ and $t_{z}$ are even in ${\bf k}$, and the spin splitting is proportional to $t_{z}$.

For concreteness, we consider a two-dimensional Hamiltonian with
$t_x = t_{x0}\cos\frac{k_x}{2}\cos\frac{k_y}{2}$
and
$t_z = t_{z0}\sin k_x\sin k_y$.
This form applies to 2D layer groups and Wyckoff positions L17(p$2_1$/b11; 2a--2b), L44(pbam; 2a--2b), and L63(p4/mbm; 2b). It is also the 2D limit ($k_z=0$) of the 3D space groups 55(Pbam; 2a--2d), 58(Pnnm; 2a--2d), 127(P4/mbm; 2c--2d), and 136(P4$_2$/mnm; 2a--2b) \cite{roig2024}. This includes several widely discussed altermagnets with $k_xk_y$ spin splitting, including $\kappa$-Cl\cite{Naka2019-th}, FeSb$_2$\cite{mazin2021}, CaCrO$_3$\cite{naka2021}, MnO$_2$\cite{noda2016}, MnF$_2$\cite{yuan2020}, and LaMnO$_3$\cite{yuan2021}. In the following calculations, we focus on a tetragonal example with
$\varepsilon_{\bf k} = t_0(\cos k_x + \cos k_y) - \mu$.

The light field enters through the Peierls substitution,
$h(t)=h({\bf k}+{\bf A}(t))$,
with
${\bf A}(t)=(A_1\cos 2\omega t,\, A_2\cos\omega t)$.
The eigenstates are expressed as Floquet-Bloch states,
$|\psi(t)\rangle=\exp(-i\varepsilon t)|\phi(t)\rangle$,
where the quasienergy $\varepsilon$ is defined modulo $\omega$.
The Floquet state is periodic,
$|\phi(t)\rangle=|\phi(t+2\pi/\omega)\rangle$,
and satisfies
$
(\varepsilon+i\partial_t)|\phi(t)\rangle=h(t)|\phi(t)\rangle$.
In frequency space, the eigenvalue equation becomes
\begin{equation}
(\varepsilon+m\omega)|\phi^{(m)}\rangle
=
\sum_{m'}h^{(m-m')}|\phi^{(m')}\rangle,
\label{E:2}
\end{equation}
where $|\phi^{(m)}\rangle$ and $h^{(m)}$ denote the $m$th Fourier components of the Floquet state and the time-periodic Hamiltonian, respectively. Here, the $A_x$ component with frequency $2\omega$ contributes only to even Fourier components of $h^{(m)}$, whereas the $A_y$ component with frequency $\omega$ contributes to all orders. The resulting Floquet Hamiltonian can then be diagonalized numerically (See Method).
\begin{figure}[h]
\centering
\includegraphics[width=0.9\linewidth]{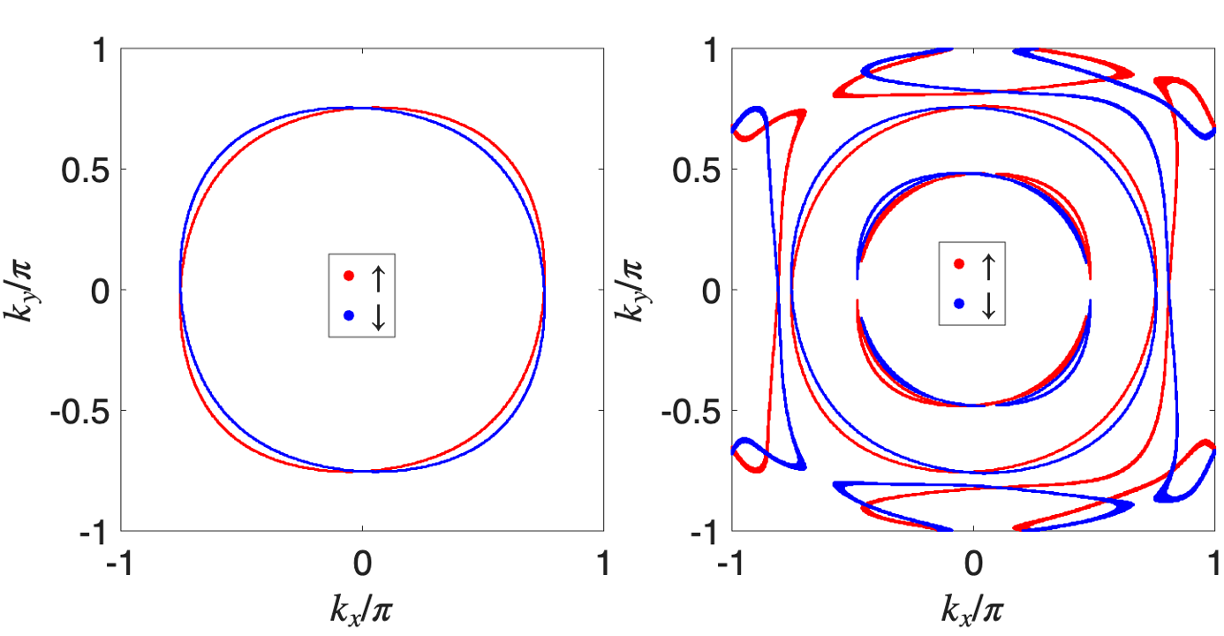}
\caption{Spin-split Fermi surfaces.
(Left panel) In the absence of light, the altermagnet exhibits even-parity
$k_xk_y\sigma_z$ spin splitting.
(Right panel) Under two-color driving, an additional odd-parity
$k_y\sigma_z$ spin splitting is induced.
Parameters (in eV): $t_0=-0.02$, $t_{x0}=0.33$, $t_{z0}=0.03$, $\mu=-0.14$,
and $J=0.05$, corresponding to the $k_z=0$ plane of MnF$_2$\cite{roig2024}.
The light parameters are $\omega=0.12$~eV and
$\sqrt{2}A_1=A_2=0.5a_0^{-1}$, with lattice constant $a_0=5$~\AA.}
\label{F:2}
\end{figure}

The Fermi surface at $\varepsilon=0$ is shown in Fig.\ref{F:2}.
In the absence of light, the altermagnet exhibits
$k_xk_y\sigma_z$ spin splitting (left panel).
Under two-color driving, an additional
$k_y\sigma_z$ spin splitting is induced (right panel), consistent with the group-theoretical analysis.
The resulting total spin splitting breaks inversion and $M_xC_{2x}^S$ symmetry while preserving $M_yC_{2x}^S$ symmetry, where $M_x$ (or $M_y$) denotes the real-space mirror reflection
$k_x\to -k_x$ (or $k_y\to -k_y$), and $C_{2x}^S$ denotes the twofold spin rotation that flips $\sigma_z$.
Band renormalization and the induced spin splitting become substantial under a light strength $A\sim O(0.1-1) a_0^{-1}$, consistent with experimentally relevant values used in Floquet topological\cite{kitagawa2011} and spintronic\cite{Zhou2025,zhu2025,li2025,huang2025,fu2026,fu2026floquet}systems.

\red{Apart from the quantitative numerical results, we provide here an analytical perturbative calculation in the limits $\omega\gg t_{x,z}$ and $A\ll 1$. This calculation offers only a qualitative understanding of the results, and the previous numerical results do not depend on this approximation. Notably, frequency $\omega$ comparable to the bandwidth is experimentally preferable because it couples strongly to the electronic states (as considered in Fig.\ref{F:2}).} 
The DC response is captured by the static ($n=0$) effective Hamiltonian:
\begin{equation}
h^{(0)}_{\rm eff} = h^{(0)}
+ \frac{[h^{(-1)},h^{(1)}]}{\omega}
+ \frac{[h^{(-2)},h^{(2)}]}{2\omega}
+ \mathcal{O}\left(\frac{1}{\omega^2}\right).
\label{E:per}
\end{equation}
Since
$h(t)=h(-t)$, Fourier components satisfy
$h^{(+m)}=h^{(-m)}$,
so the commutators vanish. This reflects the preserved $PT$ symmetry: a nonzero commutator between
$\tau_{x,z}$
would generate the $PT$-odd operator
$\tau_y$.
The leading $\mathcal{O}(A^3)$ contribution to
$h^{(0)}_{\rm eff}$ takes the form
$h_L = l_{x,{\bf k}}\tau_x + l_{z,{\bf k}}\tau_z + l_{0,{\bf k}}$, which modify the dispersion in Eq.\ref{E:1} to $E_{\alpha\beta}=\varepsilon_0+l_0+\alpha\sqrt{(t_x+l_x)^2+(t_z+l_z+\beta J)^2},
\alpha,\beta=\pm$,
so that the spin splitting is proportional to
$t_z+l_z$.
In addition to the altermagnetic contribution
$t_z\sim k_xk_y$,
the light field induces an additional spin splitting $l_z$. Explicit evaluation in Supplementary Material (SM) gives
$l_0=0$,
$l_x\propto k_x$,
and $l_z
\propto k_y$.

The induced term $h_L$ has the same form as a translationally-invariant $P$-odd loop-current order.
In general, the ${\bf k}$-odd term $l_x$ describes inter-sublattice currents,
whereas the ${\bf k}$-odd terms $l_z$ (or $l_0$) describe the difference (or average)
of intra-sublattice currents between the two sublattices. The momentum dependence of
$l_{x,z}$ determines the orientation of the current flow.
From the dispersion above, coupling a $P$-odd loop-current order to an altermagnet
in the absence of light will generally induce an odd-parity spin splitting
proportional to $l_z$.

{\it Light coupled to $PT$-symmetric magnet:}
We now discuss the external or internal $(P,T)=(-,-)$ order coupled to collinear $PT$-symmetric magnets. Since the $PT$-symmetric magnetic state has $(P,T)=(-,-)$, such coupling induces a distinct magnetic phase with $(P,T)=(+,+)$.
Importantly, $P\times T$ symmetry is preserved by both magnetic phases and the coupling, resulting in doubly degenerate bands. However, since the magnetic phase breaks spin-rotational symmetry, nontrivial spin responses can still emerge. The $(P,T)=(+,+)$ character resembles that of atomic $\vec{L}\cdot\vec{S}$ SOC, thereby allowing nonrelativistic dissipationless anomalous spin Hall conductivity.

In Landau theory, a $PT$-symmetric magnetic order $\vec{R}$ transforms as
$\Gamma_R\otimes\Gamma_A^S$. The $(P,T)=(+,+)$ magnetic state $\vec{S}$ transforms as
$\Gamma_S\otimes\Gamma_A^S
\subset
\Gamma_L\otimes\Gamma_R\otimes\Gamma_A^S$,
and is induced through the Landau coupling
$(\vec{S}\cdot\vec{R})L_A$
under two-color driving (or equivalently through
$(\vec{S}\cdot\vec{R})L_C$
when coupling to a $P$-odd loop-current order $L_C$).

DC spin conductivities are defined by
$J_i^l=\sigma_{ij}^lE_j$,
where $J_i^l$ denotes a spin current flowing along $\hat{i}$ with spin along $\hat{l}$, and $E_j$ is the electric field along $\hat{j}$. Since
$J_i^l\in\Gamma_V\otimes\Gamma_A^S$ and $E_j\in\Gamma_V\otimes\Gamma_1^S$, the spin conductivity tensor transforms as $(\Gamma_V\otimes\Gamma_V)\otimes\Gamma_A^S$.
Spin-rotational symmetry requires
$\hat{l}\parallel\vec{S},\vec{R}$,
while real-space symmetries require
$\Gamma_S\subset\Gamma_V\otimes\Gamma_V$.
Under time reversal,
$J_i^l$,
$E_j$,
and
$\vec{S}$
are all $T$-even. The DC spin conductivities
$\sigma_{ij}^l$
are thus even functions of the relaxation time $\tau$\cite{Freimuth2014,Jungwirth2017,watanabe2024,oiwa2022} and can be dissipationless.

\red{As an example, we consider the tetragonal point group $D_{4h}$, which is relevant for the metallic antiferromagnet CuMnAs\cite{wadley2015,linn2023}. The light induces $\Gamma_L\sim k_xk_y^2\in E_u$. For a PT-symmetric magnetic with $\Gamma_R\sim k_z\in A_{2u}$ , the induced $(P,T)=(+,+)$ magnetic state has $\Gamma_S=E_u\otimes A_{2u}=E_g$. This gives rise to an anomalous spin Hall conductivity $\sigma_{xz}^l=-\sigma_{zx}^l$.

Notably, the longitudinal spin conductivity $\sigma_{xz}^l=\sigma_{zx}^l$ is also symmetry-allowed. However, this contribution is forbidden by the interplay between time-reversal, spin-rotational symmetry and the equilibrium Kubo formula.} For any collinear magnet with
$\vec{R}\parallel\hat{l}$,
the spin
$\sigma_l$
is a good quantum number, and the Hamiltonian decomposes into sectors with
$\sigma_l=\pm$.
The spin conductivity
$\sigma_{ij}^{l}$
can therefore be expressed as the difference between the electric conductivities
$\sigma_{ij}$
in these two sectors. In equilibrium, the longitudinal electric conductivities
$\sigma_{ij}=\sigma_{ji}$
are odd functions of the relaxation time $\tau$, as in Drude formula. By contrast, the transverse Hall conductivity
$\sigma_{ij}=-\sigma_{ji}$ are even function of $\tau$. Since spin conductivities from $\vec{S}$ are even function of $\tau$, as dictated by time-reversal symmetry, they must be purely transverse.

This behavior contrasts with that of $(P,T)=(+,+)$ magnetic phases from coplanar unit-cell-doubling AFM order\cite{yu2026}, where
$\sigma_l$
is not a good quantum number and dissipationless longitudinal spin conductivities can emerge. In fact, the spin conductivities can be purely longitudinal when allowed by symmetry.

We perform a microscopic Floquet analysis to verify the above results.
A general two-sublattice model for a collinear $PT$-symmetric magnet reads
\begin{equation}
h = t_{x,{\bf k}}\tau_x + t_{y,{\bf k}}\tau_y + J\tau_z\sigma_z
  + \varepsilon_{0,{\bf k}},
  \label{E:PT}
\end{equation}
where $\tau$ and $\sigma$ are Pauli matrices acting in the sublattice and spin spaces, respectively. Time-reversal and inversion symmetries are represented by
$T=i\sigma_y K$
and
$P=\tau_x$
together with momentum inversion. Spin ordering is opposite on the two sublattices, interchanged by inversion. The ${\bf k}$-even hopping parameter $t_x$ and ${\bf k}$-odd hopping parameter $t_y$ are listed for various space groups and Wyckoff positions in Ref.\cite{yu2025}.

\begin{figure}[h]
\centering
\includegraphics[width=0.8\linewidth]{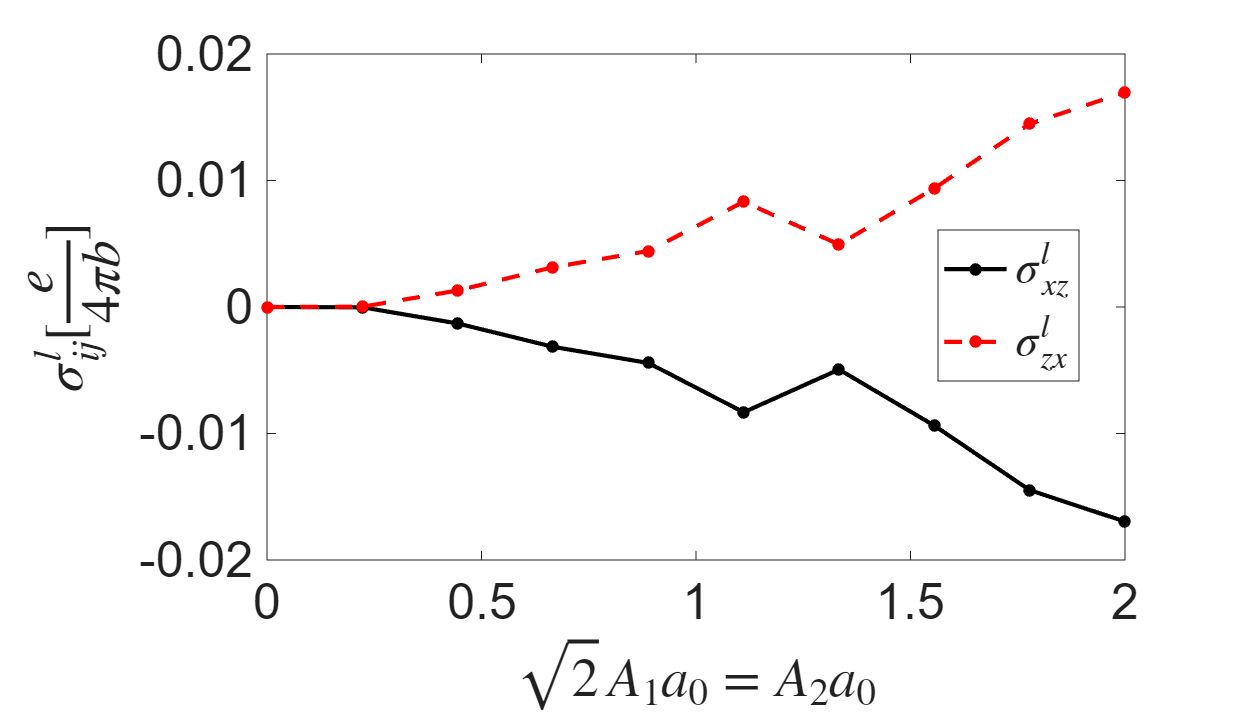}
\caption{\red{Light-induced 3D nonrelativistic dissipationless transverse spin Hall conductivities $\sigma_{xz}^l=-\sigma_{zx}^l$ for a collinear $PT$-symmetric magnet. Parameters (in unit of $t_0$): $t_{x0}=5$, $t_{y0}=0.8$, $t_z=-0.1$, $J=4.2$, $\omega=2$, and  $\mu=4.45$ are taken. The resulting Fermi surface can be found in SM. b is the lattice spacing in the y-direction.}}
\label{F:3}
\end{figure}

\red{For concreteness, we consider space group 129 (P4/nmm) for CuMnAs. For Wyckoff position 2c (magnetic atoms Mn), the hopping parameters are: $t_{x,{\bf k}}=t_{x0}\cos\tfrac{k_x}{2}\cos\tfrac{k_y}{2}$,
$t_{y,{\bf k}}=t_{y0}\cos\tfrac{k_x}{2}\cos\tfrac{k_y}{2}\sin k_z$,
and
$\varepsilon_{0,{\bf k}}=t_0(\cos k_x+\cos k_y)+t_{z}\cos k_z-\mu$.}
The spin Hall conductivity is calculated using the standard Kubo formula (see Method), and the result is in Fig.\ref{F:3}.

{\it Summary:}
We have proposed two strategies for engineering nonrelativistic spin responses in collinear altermagnets: applying two-color linearly polarized light and coupling to a $P$-odd loop-current order. Both mechanisms carry $(P,T)=(-,-)$ symmetry. When coupled to an altermagnet, they induce odd-parity spin splitting. When coupled to a $PT$-symmetric magnet, they generate a distinct $(P,T)=(+,+)$ magnetic phase, enabling nonrelativistic dissipationless anomalous spin Hall conductivity. These results are supported by both Landau theory and Floquet band-structure calculations. Our work establishes two-color light and $P$-odd loop-current order as versatile tools for engineering nonrelativistic spin-splitting symmetries and opens new avenues for spintronic applications.

We acknowledge with thanks the useful discussions with Daniel F. Agterberg, Karen Lau, Michael Weinert, Tatsuya Shishidou, Mercè Roig and Dhruv Upadhyaya.

\bibliography{citation}

\section{Method}
The eigenvalue equation Eq.~\ref{E:2} can be written in matrix form as
$H|\vec{\phi}\rangle=\varepsilon|\vec{\phi}\rangle$, with
\begin{equation}
\begin{split}
H=\left(\begin{array}{ccccc}
...&&&&...\\
&h^{(0)}-\omega&h^{(1)}&h^{(2)}&\\
&h^{(-1)}&h^{(0)}&h^{(1)}&\\
&h^{(-2)}&h^{(-1)}&h^{(0)}+\omega&\\
...&&&&...
\end{array}\right),
\end{split}
\label{E:BFH}
\end{equation}
and
$|\vec{\phi}\rangle=(...,|\phi^{(1)}\rangle,|\phi^{(0)}\rangle,|\phi^{(-1)}\rangle, ...)^T$.
For all numerical calculations, $H$ is truncated to an $84\times84$ matrix with
$|\vec{\phi}\rangle=(|\phi^{(10)}\rangle,...,|\phi^{(-10)}\rangle)$. Terms up to order $A^3$ and Fourier components up to $h^{(\pm4)}$ are considered.
In the right panel of Fig.~2, we plot the Fermi surface using the criterion
$|\varepsilon|<0.0005$~eV.
The left panel corresponds to
$A_{1,2}=0$,
where the original Fermi surface is plotted using
$|E|<0.0005$~eV.
We use a
$1000\times1000$
momentum-space grid.
For Fig.~3, we use a
$N=100\times100\times40$ grid.

After diagonalizing the Bloch-Floquet Hamiltonian $H$, we apply the Kubo formula to compute the 3D spin conductivities:
\begin{equation*}
\sigma_{xz}^l=-\frac{2e}{Nb}\text{Im}\sum_{\bf k}\sum_{m,n}\frac{f(E_m)-f(E_n)}{(E_m-E_n)^2}\langle m|\hat{J}_x|n\rangle\langle n|\hat{J}_z^l|m\rangle
\end{equation*}
The current and spin-current operators are
$\hat{J}_i=\partial H/\partial k_i$
and
$\hat{J}_j^l=\frac{1}{2}\{\partial H/\partial k_j,\sigma_l/2\}$,
respectively.
Here,
$f(E)=\theta(-E)$
is the zero-temperature Fermi-Dirac distribution. b is the lattice spacing in the y direction.

If the spins are collinear along $\hat{z}$, $|m\rangle$ can be chosen to be eigenstates of $\sigma_z$, and spin current becomes $\hat{J}_j^z=\frac{\sigma_z}{2}\partial H/\partial k_j$. The above Kubo formula can be evaluated separately in the
$\sigma_z=\pm$
sectors as
$\sigma_{ij}^z=\sigma_{ij}^+-\sigma_{ij}^-$,
where
\begin{equation*}
\begin{split}
&\sigma_{ij}^\pm=\mp\frac{e}{Nb}\text{Im}\sum_{\bf k}\sum_{m,n}\frac{f(E_m)-f(E_n)}{(E_m-E_n)^2}\langle m|\hat{J}_i|n\rangle\langle n|\hat{J}_j|m\rangle,
\end{split}
\end{equation*}
for
$|m\rangle,|n\rangle\in \sigma_z=\pm$
sectors.
Because of the factor
$f(E_m)-f(E_n)$
in the numerator, the Kubo formula requires
$\sigma_{ij}^z=-\sigma_{ji}^z$.
This applies to all DC and dissipationless spin conductivities in collinear systems, independent of symmetry requirements.
The spin Hall conductivities are related to the difference of Berry curvature
$\Omega_{xy}=\sum_{mn}\tfrac{1}{(E_m-E_n)^2}\text{Im}(\langle m|\hat{J}_x|n\rangle\langle n|\hat{J}_y|m\rangle)$
in the two spin sectors.



The Supplemental Material contains:
(1) a real-space illustration of the $P$-odd loop-current order in $D_{4h}$,
(2) the Fourier components in
$h^{(m)}$ up to $m=\pm 4$ and $A^3$
for both the tetragonal lattices, expressed in Bessel functions of the first kind,
and 
(3) reproduced Fermi surface for antiferromagnetic metal CuMnAs.

\clearpage
\appendix
\onecolumngrid
\section{P-odd loop-current order in $D_{4h}$}
\begin{figure}[h]
	\centering
	\includegraphics[width=0.5\linewidth]{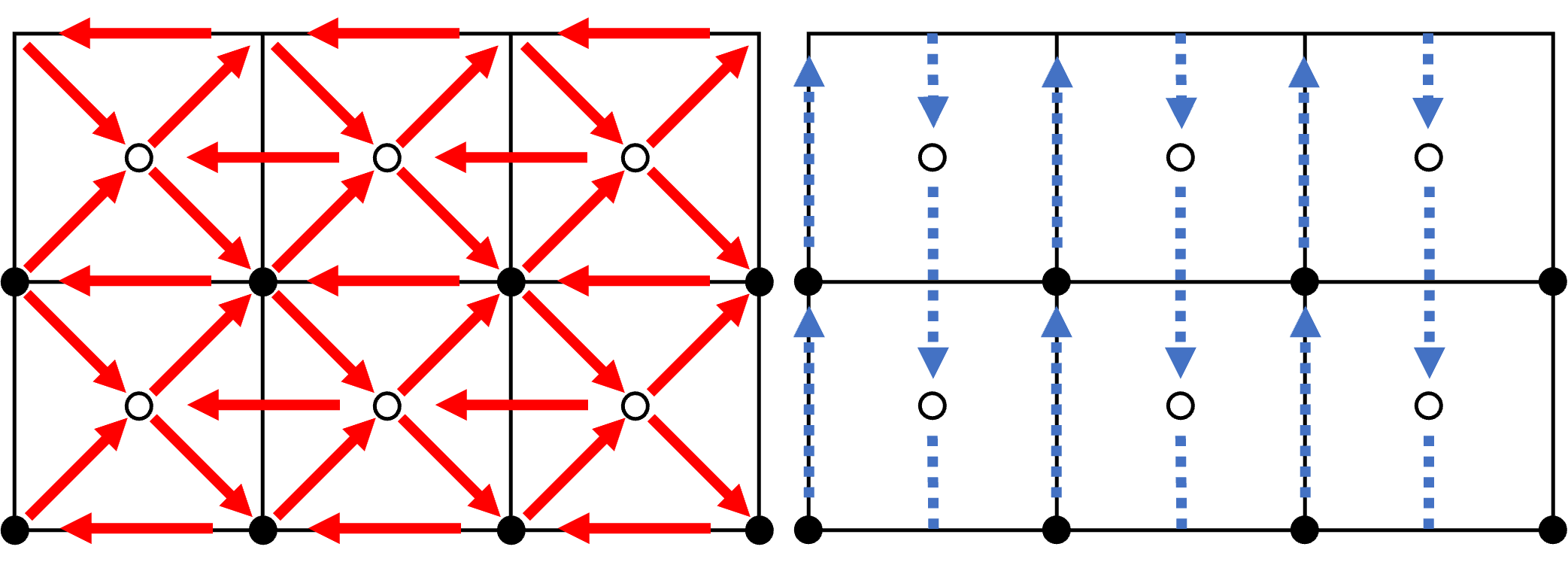}
	\caption{Translationally invariant P-odd loop-current order on bipartite square lattice. Inversion symmetry is on-site. (Left) loop-current with inter-sublattice current and net intra-sublattice current. (Right) loop-current with opposite intra-sublattice current on the two sublattices. }
	\label{F:A1}
\end{figure}

Some simplest loop-current orders in $D_{4h}$ is shown in Fig.\ref{F:A1}. As in the main text, we consider site-symmetry $B_{2g}\sim k_xk_y$. The two loop-current orders are described by $h_L=l_x\tau_x+l_z\tau_z+l_0$ with (Left) $l_x\propto\sin\frac{k_x}{2}\cos\frac{k_y}{2}$ and $l_z\propto\sin k_y$. They carry the same symmetry and generically mix. The $l_z\tau_z$ component, when combining with altermagnet $J\tau_z\sigma_z$, generates the odd-parity spin splitting $l_z\sigma_z\propto \sin k_y\sigma_z$.

\section{Fourier components in Bloch-Floquet Hamiltonian for $D_{4h}$ altermagnets}
The Fourier transformation $h^{(m)}$ can be performed using the Jacobi-Anger expansion:
$\exp(iA\cos\omega t)=\sum_{n=-\infty}^{\infty} i^nJ_n(A)\exp(in\omega t)$,
where $J_n(A)$ is the $n$th Bessel function of the first kind.
For small $A$, $J_0(A)\approx1$ and $J_n(A)\propto A^n$. In the numerical calculations, we keep terms up to order $A^3$ and Fourier components up to $h^{(\pm4)}$. We obtain hopping coefficients in $h^{(\pm m)}\equiv \epsilon_0^{(\pm m)}+t_x^{(\pm m)}\tau_x+t_z^{(\pm m)}\tau_z$ for Eq.1. For the 2D altermagnetic model $t_x = t_{x0}\cos\frac{k_x}{2}\cos\frac{k_y}{2}$, $t_z = t_{z0}\sin k_x\sin k_y$, and $\varepsilon_{\bf k} = t_0(\cos k_x + \cos k_y) - \mu$, we have:
\begin{equation}
	\begin{split}&
		\epsilon^{(0)}=t_0J_0(A_1)\cos k_x+t_0J_0(A_2)\cos k_y-\mu,
		\epsilon^{(1)}=-t_0J_1(A_2)\sin k_y\\&
		\epsilon^{(2)}=-t_0J_1(A_1)\sin k_x-t_0J_2(A_2)\cos k_y,
		\epsilon^{(3)}=+t_0J_3(A_2)\sin k_y,
		\epsilon^{(4)}=-t_0J_2(A_1)\cos k_x
	\end{split}
\end{equation}
\begin{equation}
	\begin{split}
		&t_x^{(0)}=+t_{x0}J_0(\frac{A_1}{2})J_0(\frac{A_2}{2})\cos\frac{k_x}{2}\cos\frac{k_y}{2}+2t_{x0}J_1(\frac{A_1}{2})J_2(\frac{A_2}{2})\sin\frac{k_x}{2}\cos\frac{k_y}{2},\\&
		t_x^{(1)}=-t_{x0}J_0(\frac{A_1}{2})J_1(\frac{A_2}{2})\cos\frac{k_x}{2}\sin\frac{k_y}{2}+t_{x0}J_1(\frac{A_1}{2})J_1(\frac{A_2}{2})\sin\frac{k_x}{2}\sin\frac{k_y}{2}\\&
		t_x^{(2)}=-t_{x0}J_1(\frac{A_1}{2})J_0(\frac{A_2}{2})\sin\frac{k_x}{2}\cos\frac{k_y}{2}-t_{x0}J_0(\frac{A_1}{2})J_2(\frac{A_2}{2})\cos\frac{k_x}{2}\cos\frac{k_y}{2},\\&
		t_x^{(3)}=+t_{x0}J_1(\frac{A_1}{2})J_1(\frac{A_2}{2})\sin\frac{k_x}{2}\sin\frac{k_y}{2}+t_{x0}J_0(\frac{A_1}{2})J_3(\frac{A_2}{2})\cos\frac{k_x}{2}\sin\frac{k_y}{2}+t_{x0}J_2(\frac{A_1}{2})J_1(\frac{A_2}{2})\cos\frac{k_x}{2}\sin\frac{k_y}{2}\\&
		t_x^{(4)}=-t_{x0}J_2(\frac{A_1}{2})J_0(\frac{A_2}{2})\cos\frac{k_x}{2}\cos\frac{k_y}{2}+t_{x0}J_1(\frac{A_1}{2})J_2(\frac{A_2}{2})\sin\frac{k_x}{2}\cos\frac{k_y}{2}
	\end{split}
\end{equation}
\begin{equation}
	\begin{split}
		&t_z^{(0)}=+t_{z0}J_0(A_1)J_0(A_2)\sin k_x\sin k_y-2t_{z0}J_1(A_1)J_2(A_2)\cos k_x\sin k_y,\\&
		t_z^{(1)}=+t_{z0}J_0(A_1)J_1(A_2)\sin k_x\cos k_y+t_{z0}J_1(A_1)J_1(A_2)\cos k_x\cos k_y\\&
		t_z^{(2)}=+t_{z0}J_1(A_1)J_0(A_2)\cos k_x\sin k_y-t_{z0}J_0(A_1)J_2(A_2)\sin k_x\sin k_y,\\&
		t_z^{(3)}=+t_{z0}J_1(A_1)J_1(A_2)\cos k_x\cos k_y-t_{z0}J_0(A_1)J_3(A_2)\sin k_x\cos k_y-t_{z0}J_2(A_1)J_1(A_2)\sin k_x\cos k_y\\&
		t_z^{(4)}=-t_{z0}J_2(A_1)J_0(A_2)\sin k_x\sin k_y-t_{z0}J_1(A_1)J_2(A_2)\cos k_x\sin k_y
	\end{split}
\end{equation}
Static altermagnetic terms are in $h^{(0)}$: $J\tau_z\sigma_z$. 

For perturbation theory with $\omega\gg h^{(0)}$, the leading $A^3$ contribution is in $h^{(0)}$:
\begin{equation}
	\begin{split}
		h_L&=2t_{x0}J_1(\frac{A_1}{2})J_2(\frac{A_2}{2})\sin\frac{k_x}{2}\cos\frac{k_y}{2}\tau_x-2t_{z0}J_1(A_1)J_2(A_2)\cos k_x\sin k_y\tau_z\\
		&\approx \frac{1}{64}A_1A_2^2t_{x0}\sin\frac{k_x}{2}\cos\frac{k_y}{2}\tau_x-\frac{1}{8}A_1A_2^2t_{z0}\cos k_x\sin k_y\tau_z\equiv l_{x,{\bf k}}\tau_x+l_{z,{\bf k}}\tau_z,
	\end{split}
\end{equation}
which can also be obtained from the time average $\overline{h(t)}$, as illustrated in the Method. With the k-odd coefficients $l_{x,z}$, $h_L$ carries the same form as a P-odd loop-current order. Since the system is unbounded, a net current generically coexists with the loop-current order. 

\section{Fermi surface for CuMnAs}
The two-band model produces the following 3D Fermi surfaces. This reproduces the large 3D Fermi surface reported in ARPES for CuMnAs\cite{linn2023}. Notably, DFT in Ref.\cite{linn2023} suggests another much smaller Fermi surface near the $\Gamma$ point, which can be captured if one use a 4-band model. 
\begin{figure}[h]
	\centering
	\includegraphics[width=0.45\linewidth]{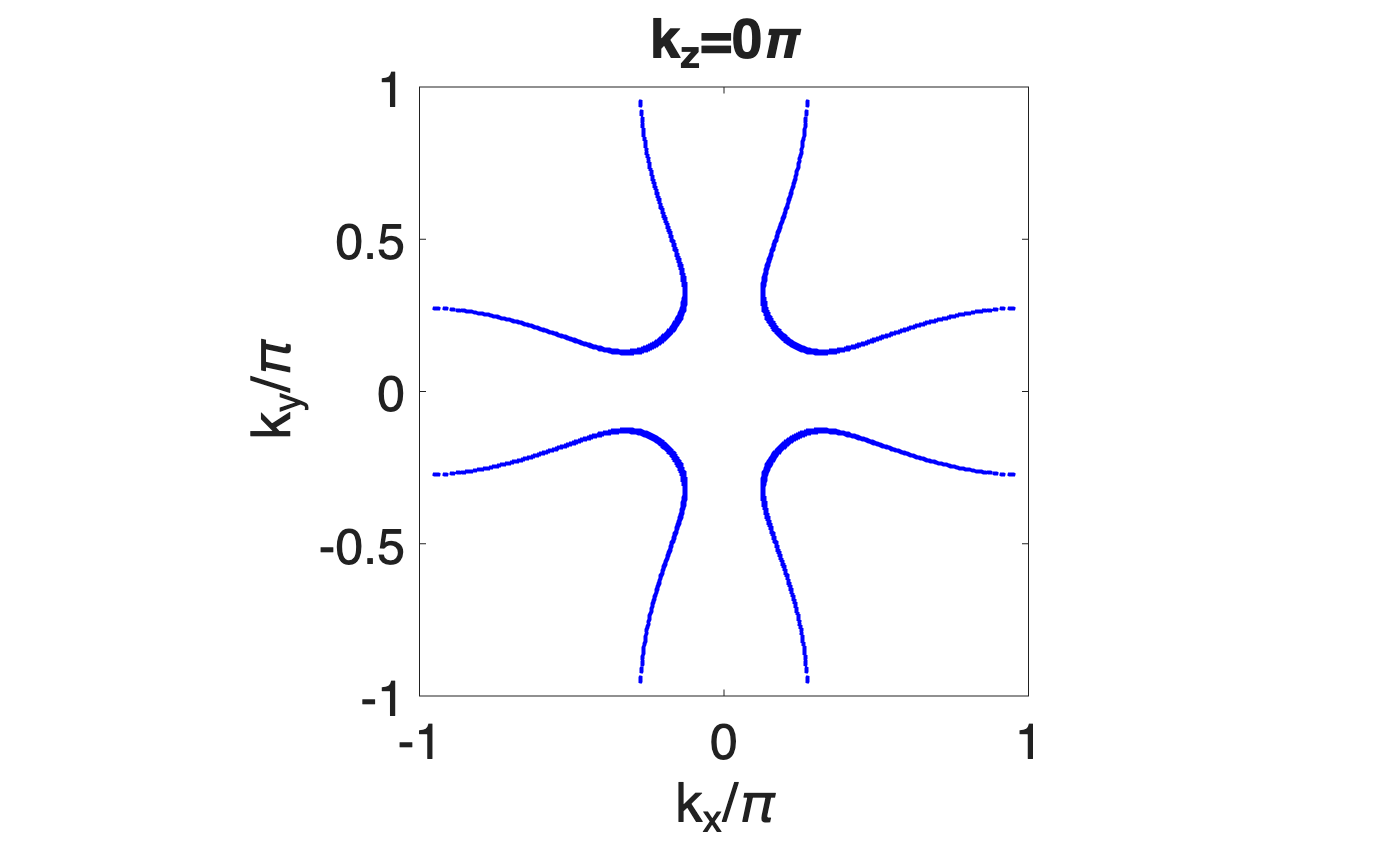}
	\includegraphics[width=0.45\linewidth]{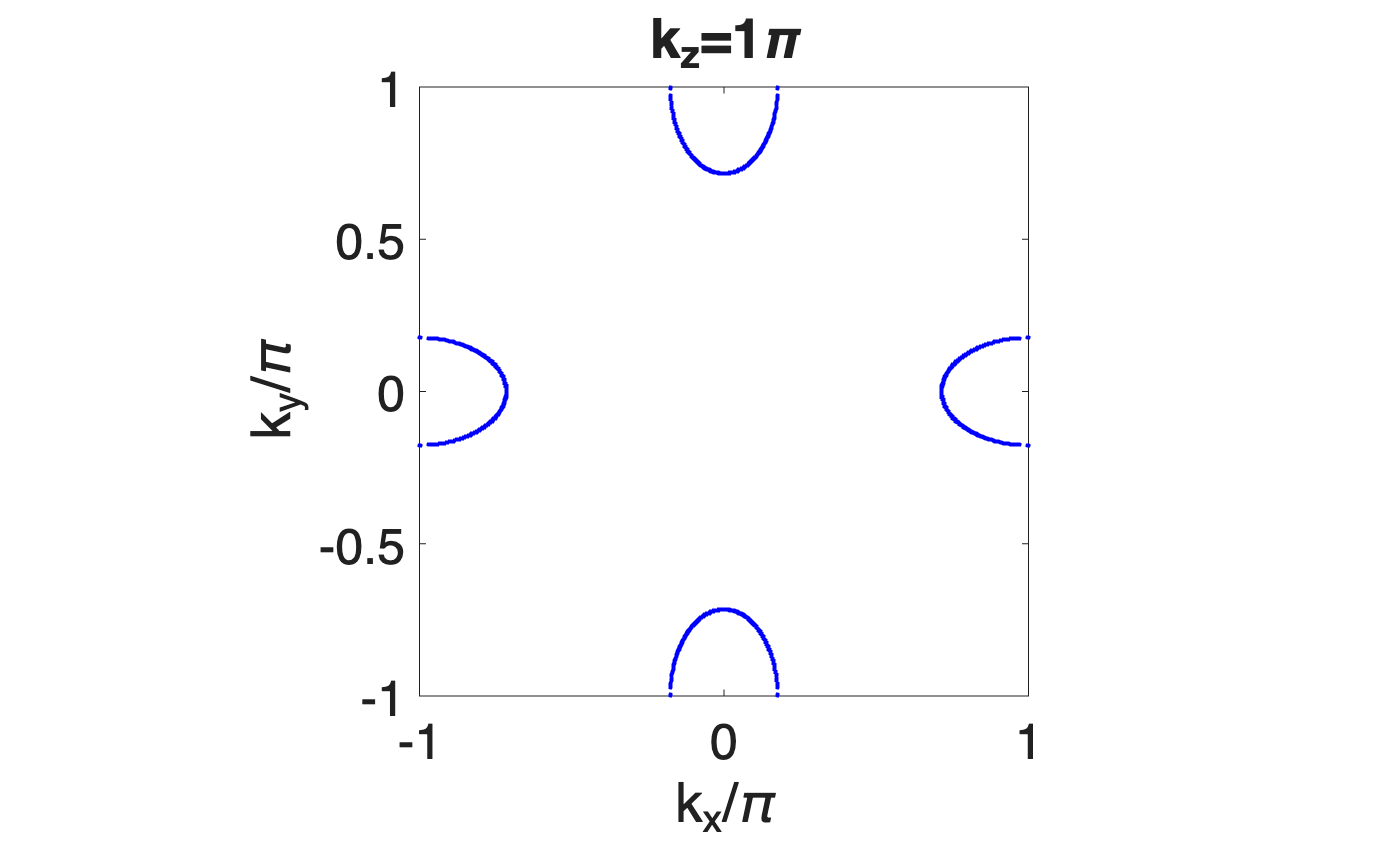}
	\caption{Fermi surfaces from two two-band tight-binding model for CuMnAs.}
\end{figure}

\end{document}